\documentclass[]{aipproc}
\layoutstyle{8d}

\newcommand{\hbp}{high-$B$ pulsars}

\newcommand{\ergs}{\,ergs\,s$^{-1}$}

\newcommand{\ppdot}{\mbox{$P$--$\dot{P}$}}
\newcommand{\pdot}{\mbox{$\dot{P}$}}
\newcommand{\edot}{\mbox{$\dot{E}$}}
\newcommand{\cxo}{\emph{Chandra}}
\newcommand{\xmm}{\emph{XMM-Newton}} 

\newcommand\apj{ApJ}%
\newcommand\apjl{ApJ}%
\newcommand\aap{A\&A}%
\newcommand\mnras{MNRAS}%
\newcommand\nat{Nature}%
\begin{document}
\title[High-$B$ Pulsars]{High Magnetic Field Rotation-powered Pulsars}
\classification{97.60.Gb,97.60.Jd}
\keywords{pulsars: general -- stars: neutron -- X-rays: stars}
\author{C.-Y.\ Ng\footnote{ncy@hep.physics.mcgill.ca; Tomlinson Postdoctoral Fellow} }{address={Department of Physics, McGill University, Montreal, QC H3A 2T8, Canada}}
\author{V.\ M.\ Kaspi}{address={Department of Physics, McGill University, Montreal, QC H3A 2T8, Canada}}

\begin{abstract}
Anomalous X-ray pulsars and soft gamma repeaters have recently emerged
as a unified class of neutron stars, identified by dramatic X-ray and
gamma-ray outbursts and via luminous X-ray pulsations, both
thought to be powered by the decay of an enormous internal magnetic
field. This ``magnetar'' hypothesis has raised the question of these
objects' physical relationship with conventional rotation-powered
pulsars (RPPs). The highest magnetic-field RPPs might therefore be expected
to be transition objects between the two populations. The recently reported
magnetar-like outburst of PSR J1846$-$0258, previously thought to be purely
rotation-powered, clearly supports this suggestion. Here we review the
observational properties of the highest magnetic-field RPPs known, and
show some common characteristics that are notable among RPPs, which are
plausibly related to their high fields. Using these objects, we consider
the evidence for proposed ``magneto-thermal evolution'' in neutron stars,
and argue that while some exists, it is not yet conclusive.  
\end{abstract}

\maketitle

\section{Introduction}
Since the discovery of radio pulsars some 40 years ago, they have
been the standard textbook examples of pulsars: fast-spinning
neutron stars converting their rotational energy into electromagnetic
radiation and particle winds. Hence, these pulsars are
also referred to as rotation-powered pulsars (RPPs). Based
on the spin period ($P$) and its time derivative (\pdot),
a number of physical parameters can be derived, including
the characteristic age
\begin{equation}
\tau_c \equiv P/2 \dot P
\end{equation}
and the spin-down luminosity 
\begin{equation}
\dot E \equiv 4\pi^2 I \dot P / P^3 \ ,
\end{equation}
where $I=10^{45}$\,g\,cm$^2$ is the assumed moment of inertia of a
neutron star. For pure dipole spin-down in vacuum, the
surface $B$-field at the magnetic equator can be estimated by
\begin{equation}
B=3.2\times 10^{19}\sqrt{P\pdot}\,G \ , \label{eqt:b}
\end{equation}
where $P$ is in s. Note that the field strength at the
magnetic poles is higher by a factor of two \citep{st83}.

\begin{figure}[ht]
\includegraphics[height=.3\textheight]{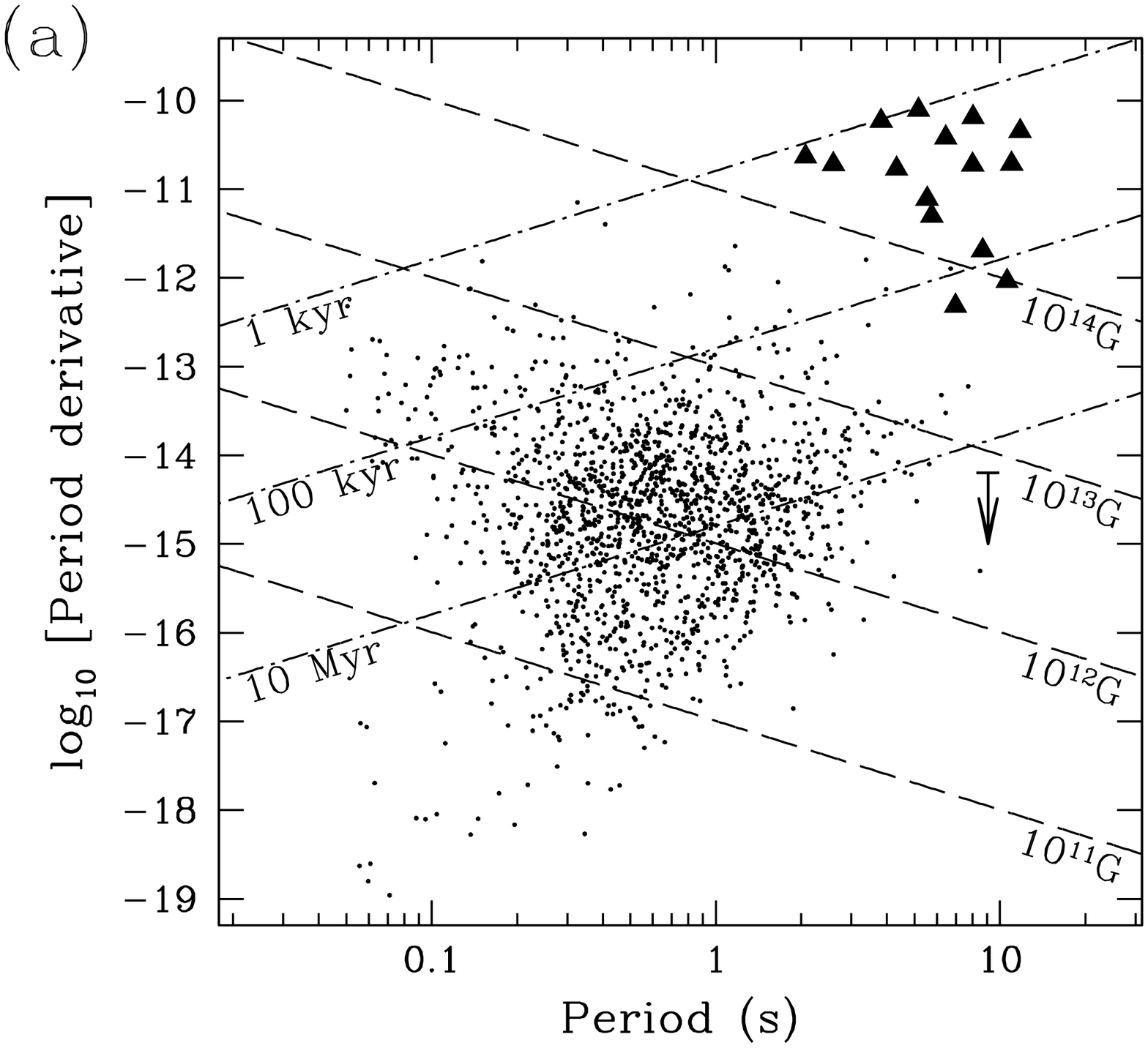}
\hspace*{10mm}
\includegraphics[height=.3\textheight]{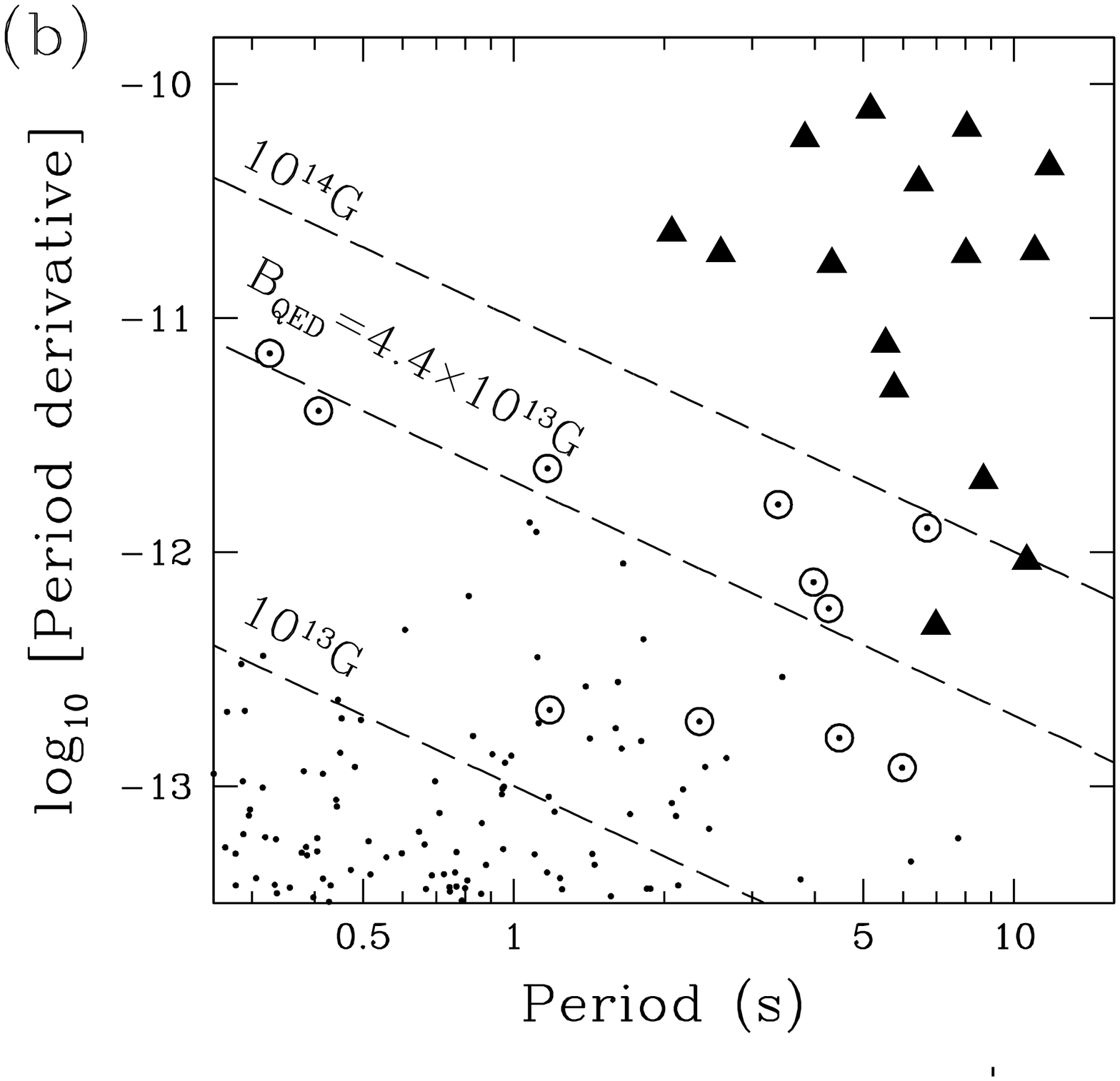}
\caption{(a) \ppdot\ diagram of isolated radio pulsars and magnetars,
represented by dots and triangles, respectively.
The upper limit on \pdot\ of the magnetar SGR 0418+5729 is
shown by the arrow \citep{ret+10}.
(b) Zoom-in of the same plot, showing the region
containing \hbp\ and magnetars. Objects listed in
Table~\ref{table} are marked by the circles.
\label{ppdot}}
\end{figure}

RPPs are characterized by their pulsations from radio to MeV
$\gamma$-rays, with a total radiation power generally
less than 1\% of \edot. Young RPPs are often associated
with pulsar wind nebulae (PWNe), which provide a unique
signature of the pulsar nature even if the radio beams miss
the Earth. Figure~\ref{ppdot}a shows the \ppdot\ diagram
of all isolated RPPs, indicating a typical $B$-field 
around $10^{12}$\,G inferred from the spin parameters.

Over the past two decades, several new classes of neutron
stars have been discovered. The most exotic one is magnetars,
a small group\footnote{A catalog of magnetars
can be found at \url{http://www.physics.mcgill.ca/~pulsar/magnetar/main.html}}
of isolated X-ray pulsars with long spin periods
and large $\dot{P}$s that imply ultra-strong surface fields of
$10^{14}-10^{15}$\,G
(see review by E.\ G{\"o}{\u g}{\"u}{\c s} in this Volume).
This field strength is well above the so-called ``quantum critical
field" of 
\begin{equation}
B_{\rm QED} \equiv m_e^2c^3/e\hbar \simeq 4.4\times 10^{13}\,\mathrm{G} \ ,
\end{equation}
at which the electron cyclotron energy is equal to its rest mass.
Some theories predict that under such a strong field, pair creation
will become ineffective due to photon splitting, thus, suppressing
the radio emission \citep{bh98,bh01}.

Magnetars were historically identified as anomalous X-ray
pulsars (AXPs) or soft gamma repeaters (SGRs) according to how
they were first discovered. In contrast to RPPs, magnetars in
an active state could have X-ray luminosities 1-2 orders
of magnitude higher than \edot. This requires an additional energy
source other than rotation, which is generally believed to be
the decay of their strong magnetic fields. The most remarkable
feature of magnetars is their violent outbursts, during which the
X-ray luminosity can increase by a few orders of magnitude.
These are often accompanied with timing anomalies
\citep[e.g.][]{kgw+03,dkg08}. While radio
pulsations have been detected from three magnetars
\citep{crh+06,crh+07,lbb+10}, their radio emission is
largely distinct from that of RPPs, including highly variable
radio flux densities, and flat or inverted
radio spectra \citep{lbb+10,crp+07,crj+08,ljk+08}. These suggest
that they could have a different radio emission mechanism than
that of the RPPs. For the purpose of this review, we do not consider these
radio-emitting magnetars as `radio pulsars', and reserve the term
for RPPs, even though some RPPs have no radio detection.

Thanks to the Parkes Multibeam Pulsar Survey (PMPS) \citep[][and
references therein]{hfs+04},
over 700 new radio pulsars were discovered and a handful
of them have spin parameters similar to those of the magnetars,
implying comparable field strengths. Table~\ref{table}
lists all known high-magnetic-field RPPs (hereafter, high-$B$
pulsars) with $B>B_{\rm QED}$, together with a few other \hbp\
that have previously been studied. The objects
are plotted in the \ppdot\ diagram in Figure~\ref{ppdot}b,
clearly indicating an overlapping parameter space with some
magnetars. Therefore, \hbp\ present an important link between
RPPs and magnetars and could help understand magnetar physics. In particular,
one might expect \hbp\ to have higher X-ray luminosities than other
radio pulsars, and possibly exhibit magnetar-like properties.
We will describe some individual sources in the next section,
and then discuss the connection of \hbp\ to other classes of neutron
stars.

\begin{table}[ht]
\begin{tabular}{cccccccc}
\hline
\tablehead{1}{c}{c}{Name\tablenote{the superscript $^{\rm R}$ indicates that the
pulsar is classified as a rotating radio transient (RRAT).}} &
\tablehead{1}{c}{c}{$P$ (s)} &
\tablehead{1}{c}{c}{$B$ ($10^{13}$\,G)} &
\tablehead{1}{c}{c}{\edot\ (\ergs)} &
\tablehead{1}{c}{c}{$\tau_c$ (kyr)} &
\tablehead{1}{c}{c}{$d$\tablenote{estimated from the dispersion measure of the
pulsars, except for PSRs J1846$-$0258 and J1119$-$6127, for which the distances are
obtained from H{\sc I} absorption measurements \citep{lt08,cmc04}.} (kpc)} &
\tablehead{1}{c}{c}{$L_X$\footnote{converted into the 0.5-10\,keV band.} (\ergs)} &
\tablehead{1}{c}{c}{Ref.} \\
\hline
J1847$-$0130 & 6.71 & 9.4 & $1.7\times10^{32}$ & 83 & 8.4 & $<3.4\times10^{34}$ & \citep{msk+03} \\
J1718$-$3718 & 3.38 & 7.4 & $1.6\times10^{33}$ & 34 & 4.5 & 0.14-2.6$\times10^{33}$ & \citep{km05} \\
J1814$-$1744 & 3.98 & 5.5 & $4.7\times10^{32}$ & 85 & 10 & $<4.3\times10^{34}$ & \citep{pkc00} \\
J1734$-$3333 & 1.17 & 5.2 & $5.6\times10^{34}$ & 8.1 & 6.1 & 0.1-3.4$\times10^{33}$ & \citep{okl+10} \\
J1819$-$1458$^{\rm R}$ & 4.26 & 5.0 & $2.9\times10^{32}$ & 117 & 3.6 & 1.8-2.4$\times10^{33}$ & \citep{mrg+07} \\
J1846$-$0258 & 0.33 & 4.9 & $8.1\times10^{36}$ & 0.9 & 6.0 & 2.5-2.8$\times10^{34}$\footnote{in 2000, prior to the 2006 outburst.} & \citep{nsg+08} \\
 & & & & & & 1.2-1.7$\times10^{35}$\footnote{during the outburst in 2006.} & \citep{nsg+08} \\
J1119$-$6127 & 0.41 & 4.1 & $2.3\times10^{36}$ & 1.7 & 8.4 & 1.9-3.5$\times10^{33}$ & \citep{sk08} \\
J0847$-$4316$^{\rm R}$ & 5.98 & 2.7 & $2.2\times10^{31}$ & 790 & 3.4 & $<6\times10^{31}$ & \citep{kec+09} \\
J1846$-$0257$^{\rm R}$ & 4.48 & 2.7 & $7.1\times10^{31}$ & 442 & 5.2 & $<1.8\times10^{32}$ & \citep{kec+09} \\
B0154+61 & 2.35 & 2.1 & $5.7\times10^{32}$ & 197 & 1.7 & $<1.4\times10^{31}$ & \citep{gkl+04} \\
B1916+14 & 1.18 & 1.6 & $5.1\times10^{33}$ & 88 & 2.1 & 1.1-2.3$\times10^{31}$ & \citep{zkg+09} \\
\hline
\end{tabular}
\caption{Measured and derived properties of \hbp}
\label{table}
\end{table}

\section{Individual High-$B$ RPPs}
In this section we describe details of some individual
high-$B$.  

\begin{itemize}
\item {\bf J1847$-$0130:}
This 6.7s-period radio pulsar was discovered in the PMPS \citep{msk+03}
and it has by far the highest known $B$-field among all RPPs
in the ATNF Pulsar Catalog\footnote{\url{http://www.atnf.csiro.au/research/pulsar/psrcat/}}.
The spin-down-inferred $B$-field is
$9.4\times10^{13}$\,G, well above the quantum critical limit,
and even higher than that of two magnetars, AXP 1E~2259+586
and SGR~0418+5729 \citep{ret+10}. This discovery demonstrates that
the pulsar radio emission
mechanism can work in such a strong field, presenting a
challenge to some theories \citep[e.g.][]{bh98}. The source
is not detected in X-rays, however.  A flux limit of
$5\times 10^{33}$\ergs\ $\approx10\edot$ (2-10\,keV) was 
obtained from \emph{ASCA} observations and is not very constraining
\citep{msk+03}.

\item {\bf J1718$-$3718:}
Also discovered in the PMPS \citep{hfs+04}, this is the second
highest $B$-field RPP. An X-ray counterpart was found serendipitously
in a \cxo\ exposure of a nearby source \citep{km05}. Deeper
follow-up \cxo\ observations have detected pulsations in the
soft X-ray band (0.8-2\,keV) and better constrain the source
spectrum \citep{zkm+10}. However, the data indicate no evidence
of long-term X-ray variability, although the pulsar exhibited
a large glitch some time between 2006 August and 2009 January
(Manchester \& Hobbs, in preparation).

\item {\bf J1734$-$3333:}
This pulsar has a strong $B$-field of $5.2\times10^{13}$\,G
inferred from the spin-down, also exceeding the QED critical limit.
Recently, \citet{elk+10} reported a braking index\footnote{The
braking index is defined as $n\equiv\nu\ddot\nu/{\dot\nu}^2$, where
$\nu$, $\dot\nu$ and $\ddot\nu$ are the spin frequency, its
time derivative and second derivative, respectively.} $n=1.0\pm0.3$
using 12 years of phase-coherent radio timing data. This could
imply a magnetic field that is growing with time, such that the trajectory
on the \ppdot\ diagram points toward the magnetar region. Therefore,
the authors suggested that this radio pulsar may be a magnetar
progenitor. Deep \xmm\ observations have identified a faint X-ray
counterpart, but found no sign of magnetar-like activity
\citep{okl+10}. The X-ray luminosity in 0.5-2\,keV is below
$0.1\dot{E}$, similar to that of a typical RPP.

\item {\bf J1819$-$1458:}
Three \hbp\ listed in Table~\ref{table} belong
to the so-called ``rotating radio transients'' (RRATs) class
of neutron stars, which are sporadic radio pulse emitters
(see M.\ McLaughlin's review in this Volume). There are nearly
50 known RRATs\footnote{\url{http://www.as.wvu.edu/~pulsar/rratalog/}},
among which a handful have spin-down measurements that
imply $B$-fields ranging from $3\times10^{12}$ to $5\times10^{13}$\,G.
Thus, not every RRAT is a high-$B$ pulsar and
the connection between these two classes of neutron stars remains
unclear. One particularly interesting object is RRAT J1819$-$1458,
which was discovered in a search for isolated bursts in the PMPS
data \citep{mll+06}.  It has the highest $B$-field of $5.0\times
10^{13}$\,G among all known RRATs \citep{lmk+09}. X-ray observations
with \xmm\ found a possible spectral feature and indicate a high
X-ray luminosity of the source, an order of magnitude larger than
\edot\ \citep{mrg+07}, suggesting that it is not entirely
rotation-powered. On the other hand, deep \cxo\ observations reveal
extended X-ray emission that could be an associated PWN
\citep{rmg+09}, a feature commonly observed among energetic RPPs.
Further studies are needed to identify the exact nature of this object.

\item {\bf J1846$-$0258:}
Located at the center of supernova remnant Kes~75 (G29.7$-$0.3),
this remarkable high-$B$ ($4.9\times10^{13}$\,G) pulsar is
one of the youngest ($\sim900$\,yr) known pulsar in our
Galaxy. Although no radio emission is detected \citep{akl+08},
it powers a bright PWN \citep{nsg+08,hcg03} and spins
down relatively steadily. This has allowed a braking index measurement
of $n=2.65\pm0.01$ \citep{lkg+06}. Therefore, this object has
long been considered as a RPP. Surprisingly, this pulsar
exhibited magnetar-like bursts in 2006 May, with
a substantial flux enhancement and spectral softening in X-rays
\citep{nsg+08,ggg+08,ks08}. Accompanied with this event was
a sizable rotational glitch followed by unusually large recovery
\citep{kh09,lkg10}. Post-outburst observations reveal an apparent
decrease in braking index of $n=2.16\pm0.13$ and larger timing noise
than before, although the X-ray emission returned
to its quiescent level months after outburst \citep{lnk+10}.
The spectacular outburst clearly indicates PSR J1846$-$0258
is a transition object between a RPP and a magnetar, raising
the possibility that some \hbp\ could be quiescent magnetars, as
first speculated by \citet{km05}.

\item {\bf J1119$-$6127:}
This young ($\tau_c = 1.7$\,kyr) and energetic ($\edot=2.3
\times10^{36}$\ergs) pulsar is associated with the supernova
remnant G292.2$-$0.5 and has a strong $B$-field of $4.1\times
10^{13}$\,G \citep{ckl+00}. The recent detection of
$\gamma$-ray pulsations with the \emph{Fermi Gamma-ray Space
Telescope} makes it the highest $B$-field $\gamma$-ray pulsar
(P.\ den Hartog's talk in this conference; Parent et al.\ in
preparation). In X-rays, it
is highly pulsed in the soft band (0.5-2\,keV) with a pulsed
fraction $74\%\pm14\%$, suggesting intrinsic anisotropy of the
thermal emission from the surface \citep{sk08,gkc+05}. With
superb spatial resolution, \cxo\ observations led to the
discovery of a faint PWN surrounding the pulsar \citep{gs03},
and helped to isolate the pulsar flux from the PWN. The pulsar
spectrum consists of thermal and non-thermal components, and
the former can be fitted by either a blackbody of temperature
$kT\sim 0.21$\,keV or a neutron star atmosphere model with
$kT\sim 0.14$\,keV \citep{sk08,gkc+05}. Hence, this object
is the youngest RPP with thermal emission detected, and also
one of the hottest. In the radio band, this pulsar exhibits
different types of behavior and shows ``RRAT-like'' emission
following glitches, possibly related to a reconfiguration of
the magnetic field \citep{wje10}.

\end{itemize}

\section{The Class of High-$B$ Pulsars}

\subsection{Connection with Magnetars}
As described above, most \hbp\ are very faint compared
to their spin-down luminosities and show no magnetar
behavior (except PSR J1846$-$0258), clearly distinct
from active magnetars. Although based purely on their
X-ray spectra, one cannot rule out the possibility
that some of the \hbp\ could be quiescent magnetars,
the three known radio-emitting magnetars show very
different radio properties than RPPs, somewhat weakening
this argument. This raise an important question: what
is the intrinsic difference between these two classes
of objects?

In Figure~\ref{ppdot} \hbp\ and magnetars occupy an
overlapping region in the \ppdot\ diagram, implying that
the spin and spin-down rate are not sufficient parameters
to determine the pulsar properties. One idea is that there
could be some ``hidden parameters'' that differentiate
the two populations, such as the neutron star
mass \citep{km05} or the $B$-field configuration
\citep{msk+03,gkw02}. In the latter picture, a magnetar
field has additional quadrupole or higher multipole
components, which have no effect on the spin-down torque.
Another attempt to unify these objects is by different
orientations of the magnetic axes with respect to the
rotation axes \citep{zh00}. This model predicts an upper
limit of $2\times10^{14}$\,G on the surface magnetic field
of a radio pulsar.

As an alternative, it is also possible that the $B$-field
inferred from spin-down is not a reliable estimator of
the true field strength due to extra spin-down torques.
One plausible scenario is spin-down under the combined
effects of magnetic braking and relativistic particle
winds \citep{hck99}. Depending on the wind luminosity,
the latter term could be substantial, resulting in
an overestimate of the surface field by an order of
magnitude if Equation~\ref{eqt:b} is assumed.  However, it
has been argued that for magnetars, their wind flows may be
episodic with small duty cycles, rendering the dipole spin-down
approximation is less biased \citep{hck99}. Another
source of spin-down could be propeller torque from a
fallback accretion disk \citep{aay01}, such that a
high-$B$ pulsar has a true surface field of only $10^
{12}$-$10^{13}$\,G, similar to those of other radio pulsars. However
in this case, it is unclear if there would be radio
emission from the pulsar, except under specific
conditions \citep[see][]{ea05}.

\subsection{Connection with Other Radio Pulsars}
\begin{figure}
  \resizebox{.95\columnwidth}{!}
   {\includegraphics{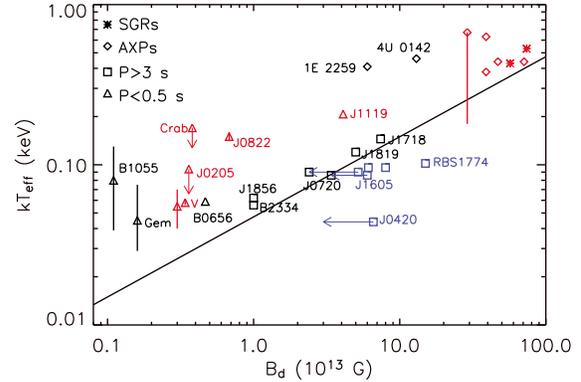}}
  \caption{Effective temperatures versus $B$-fields for
   different classes of neutron stars, adopted from
   \citet{plm+07}.\label{kt-b}}
\source{\citet{plm+07}}
\end{figure}

\citet{plm+07} first noticed an apparent correlation between
the effective temperature $T_{\rm eff}$ and surface $B$-field
in a wide range of neutron stars, with $T_{\rm eff}\propto
\sqrt{B}$ over three orders of magnitude (Figure~\ref{kt-b}).
This has motivated a series of studies on the magneto-thermal
evolution of neutron stars \citep{plm+07,apm08a,apm08b,pmg09}.
In this model, the decay of the magnetic field provides
crustal heating on the neutron star, which in turn affects
the magnetic diffusivity and thermal conductivity \citep{gkp04,
gkp06}. As a result, stars born with stronger magnetic fields
($>5\times10^{13}$\,G) are expected to show significant
field decay, which keeps them hotter for longer. While this
model is capable of explaining the relatively high X-ray temperatures of
magnetars compared with typical radio pulsars, we note that an
updated plot using recent observations of \hbp\ shows a large
scatter and the correlation seems weaker \citep{zkg+09}.

\begin{figure}
  \resizebox{.93\columnwidth}{!}
   {\includegraphics{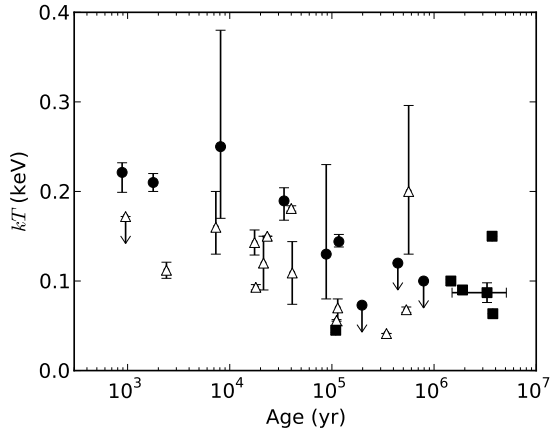}}
  \caption{Blackbody temperatures versus ages for different
  neutron stars, adopted from \citet{zkm+10}. The circles,
  triangles and squares indicate \hbp, normal radio pulsars
  and thermal isolated neutron stars, respectively.\label{kt-age}}
\source{\citet{zkm+10}}
\end{figure}

Another important parameter to consider is the pulsar age.
Since the $B$-field and temperature likely decay at
different rates, the plot above, which contains an ensemble
of pulsars at different ages, could be biased. A comparison
between the pulsar temperature and characteristic age
indicates that \hbp\ appear to be systematically hotter than other
radio pulsars (Figure~\ref{kt-age}), providing some support
to the crustal heating model \citep[][]{zkg+09,zkm+10,apm08a}.
However, the data quality does not allow one to rule out
the minimal cooling scenario. Thus, it remains unclear
if $B$-field decay is a significant source of heating
for \hbp\ \citep{zkm+10}.

\begin{figure}
  \resizebox{.99\columnwidth}{!}
   {\includegraphics{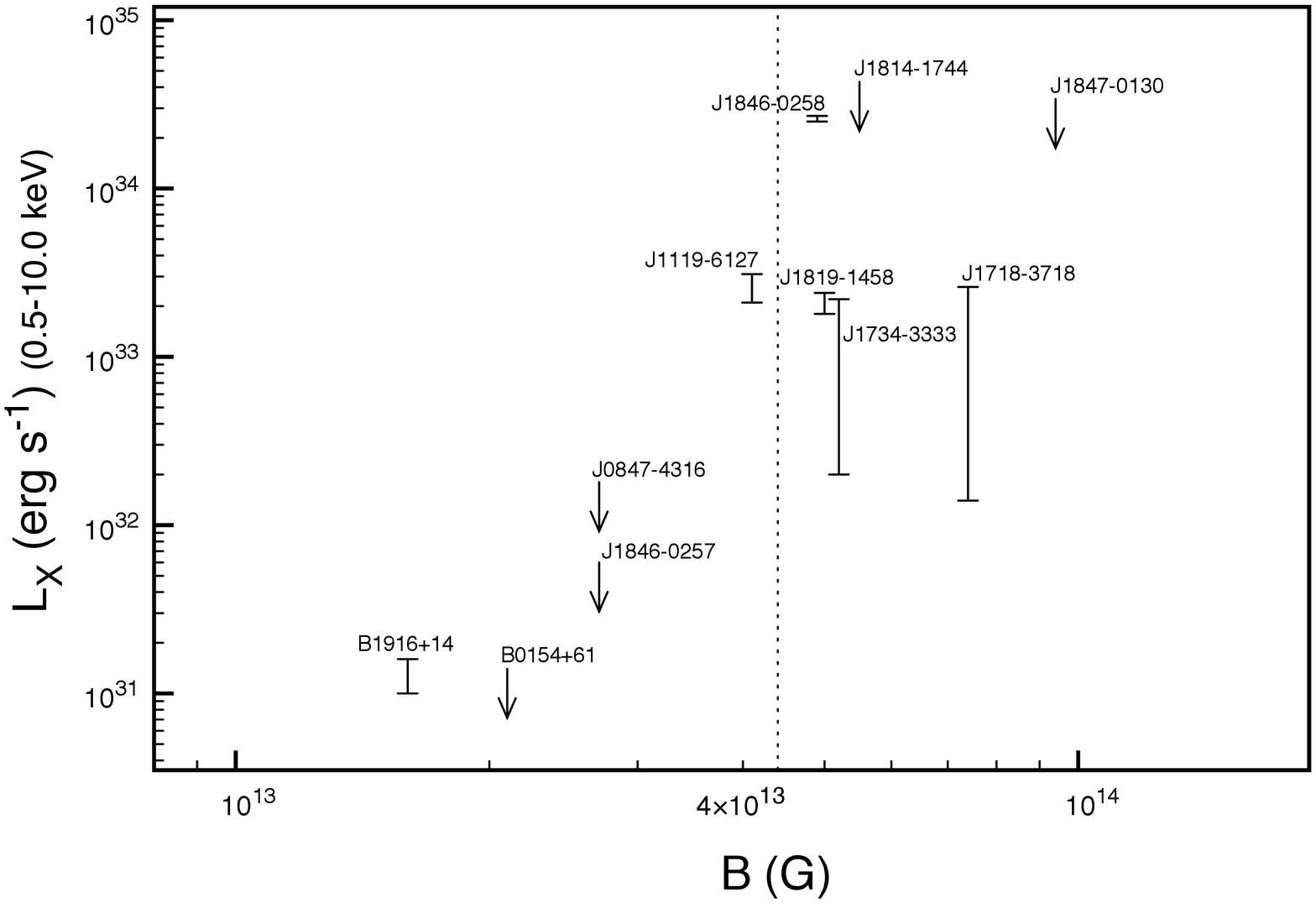}}
  \caption{X-ray luminosities versus $B$-fields for \hbp,
  adopted from \citet{okl+10}. \label{lx-b}}
\source{\citet{okl+10}}
\end{figure}

Adding to the complication is that temperature measurements
depend sensitively on the detailed physics of neutron star
atmosphere, which is not fully understood. The X-ray
luminosity, on the other hand, is less model-dependent, and
it could offer a more robust comparison between different
classes of objects. Figure~\ref{lx-b} plots the X-ray
luminosities against field strengths for the \hbp\ listed in
Table~\ref{table}, showing a hint of correlation
\citep{okl+10}. However, many pulsars in the plot are not detected in X-rays
or their luminosities are poorly constrained. Also, four out
of five brightest objects are also the youngest, hence
the possible trend may merely reflect luminosity evolution with time
(except, interestingly, PSR J1819$-$1458 which has a relatively
large characteristic age; see Table~\ref{table}).

\section{Future Prospects}
The case of PSR J1846$-$0258 has provided an important link between
the classes of RPPs and magnetars. For further study, it is crucial
to obtain more examples of transition objects. While detecting
magnetar-like outbursts would give unambiguous evidence of
such an object, it is observationally challenging because these
events could be much less energetic than those in magnetars.
In the absence of sensitive all-sky X-ray monitors, any timing
anomaly may be a good indicator of a radiative event, as is
often seen for magnetars. In practice, this would require
regular radio timing observations with prompt X-ray follow-up
after a glitch. Regular X-ray monitoring of \hbp\ is also
useful for identifying long-term flux variability, which is
another distinct feature of magnetars \citep[e.g.][]{dkg09}.

Nearly half of the \hbp\ listed in Table~\ref{table}
have only upper limits on their X-ray flux. Completing
the sample will require deep X-ray observations, ideally
with next-generation telescopes that have a
large collecting area, such as the \emph{International
X-ray Observatory (IXO)}. This will give high quality
spectra to pin down the surface temperature of \hbp,
and potentially reveal any spectral features that could
provide direct measurements of the surface field strength.

\begin{theacknowledgments}
We thank W.\ Zhu for useful discussion and for providing Figure~\ref{kt-age}.
We thank S.\ Olausen for providing Table~\ref{table} and Figure~\ref{lx-b}.
CYN is a CRAQ postdoctoral fellow.  VMK holds holds a Canada Research Chair
and the Lorne Trottier Chair, and acknowledges support from NSERC, CIFAR, and FQRNT
via CRAQ.
\end{theacknowledgments}

\bibliographystyle{aipproc}

\begin{thebibliography}{54}
\expandafter\ifx\csname natexlab\endcsname\relax\def\natexlab#1{#1}\fi
\providecommand{\enquote}[1]{``#1''}
\expandafter\ifx\csname url\endcsname\relax
  \def\url#1{\texttt{#1}}\fi
\expandafter\ifx\csname urlprefix\endcsname\relax\def\urlprefix{URL }\fi
\providecommand{\eprint}[2][]{\url{#2}}

\bibitem[{Shapiro} and {Teukolsky}(1983)]{st83}
S.~L. {Shapiro}, and S.~A. {Teukolsky}, \emph{{Black holes, white dwarfs, and
  neutron stars: The physics of compact objects}}, New York: Wiley
  Interscience, 1983.

\bibitem[Rea et al.(2010)]{ret+10} N. {Rea}, et al.\ 
\emph{Science}  (2010), in press, \eprint{arXiv:1010.2781}.

\bibitem[{Baring} and {Harding}(1998)]{bh98}
M.~G. {Baring}, and A.~K. {Harding}, \emph{\apjl} \textbf{507}, L55
  (1998).

\bibitem[{Baring} and {Harding}(2001)]{bh01}
M.~G. {Baring}, and A.~K. {Harding}, \emph{\apj} \textbf{547}, 929 (2001).

\bibitem[{Kaspi} et~al.(2003)]{kgw+03}
V.~M. {Kaspi}, F.~P. {Gavriil}, P.~M. {Woods}, J.~B. {Jensen}, M.~S.~E.
  {Roberts}, and D.~{Chakrabarty}, \emph{\apjl} \textbf{588}, L93 (2003).

\bibitem[{Dib} et~al.(2008)]{dkg08}
R.~{Dib}, V.~M. {Kaspi}, and F.~P. {Gavriil}, \enquote{{Glitches in Anomalous
  X-ray Pulsars},} in \emph{40 Years of Pulsars: Millisecond Pulsars, Magnetars
  and More}, edited by {C.~Bassa, Z.~Wang, A.~Cumming, \& V.~M.~Kaspi}, 2008,
  vol. 983 of \emph{American Institute of Physics Conference Series}, p. 239.

\bibitem[{Camilo} et~al.(2006)]{crh+06}
F.~{Camilo}, S.~M. {Ransom}, J.~P. {Halpern}, J.~{Reynolds}, D.~J. {Helfand},
  N.~{Zimmerman}, and J.~{Sarkissian}, \emph{\nat} \textbf{442}, 892
  (2006).

\bibitem[{Camilo} et~al.(2007{\natexlab{a}})]{crh+07}
F.~{Camilo}, S.~M. {Ransom}, J.~P. {Halpern}, and J.~{Reynolds}, \emph{\apjl}
  \textbf{666}, L93 (2007{\natexlab{a}}).

\bibitem[{Levin} et~al.(2010)]{lbb+10}
L.~{Levin}, M.~{Bailes}, S.~{Bates}, N.~D.~R. {Bhat}, M.~{Burgay},
  S.~{Burke-Spolaor}, N.~{D'Amico}, S.~{Johnston}, M.~{Keith}, M.~{Kramer},
  S.~{Milia}, A.~{Possenti}, N.~{Rea}, B.~{Stappers}, and W.~{van Straten},
  \emph{\apjl} \textbf{721}, L33 (2010).

\bibitem[{Camilo} et~al.(2007{\natexlab{b}})]{crp+07}
F.~{Camilo}, S.~M. {Ransom}, J.~{Pe{\~n}alver}, A.~{Karastergiou}, M.~H. {van
  Kerkwijk}, M.~{Durant}, J.~P. {Halpern}, J.~{Reynolds}, C.~{Thum}, D.~J.
  {Helfand}, N.~{Zimmerman}, and I.~{Cognard}, \emph{\apj} \textbf{669},
  561 (2007{\natexlab{b}}).

\bibitem[{Camilo} et~al.(2008)]{crj+08}
F.~{Camilo}, J.~{Reynolds}, S.~{Johnston}, J.~P. {Halpern}, and S.~M. {Ransom},
  \emph{\apj} \textbf{679}, 681 (2008).

\bibitem[{Lazaridis} et~al.(2008)]{ljk+08}
K.~{Lazaridis}, A.~{Jessner}, M.~{Kramer}, B.~W. {Stappers}, A.~G. {Lyne},
  C.~A. {Jordan}, M.~{Serylak}, and J.~A. {Zensus}, \emph{\mnras} \textbf{390},
  839 (2008).

\bibitem[{Hobbs} et~al.(2004)]{hfs+04}
G.~{Hobbs}, A.~{Faulkner}, I.~H. {Stairs}, F.~{Camilo}, R.~N. {Manchester},
  A.~G. {Lyne}, M.~{Kramer}, N.~{D'Amico}, V.~M. {Kaspi}, A.~{Possenti}, M.~A.
  {McLaughlin}, D.~R. {Lorimer}, M.~{Burgay}, B.~C. {Joshi}, and F.~{Crawford},
  \emph{\mnras} \textbf{352}, 1439 (2004).

\bibitem[{McLaughlin} et~al.(2003)]{msk+03}
M.~A. {McLaughlin}, I.~H. {Stairs}, V.~M. {Kaspi}, D.~R. {Lorimer},
  M.~{Kramer}, A.~G. {Lyne}, R.~N. {Manchester}, F.~{Camilo}, G.~{Hobbs},
  A.~{Possenti}, N.~{D'Amico}, and A.~J. {Faulkner}, \emph{\apjl} \textbf{591},
  L135 (2003).

\bibitem[{Kaspi} and {McLaughlin}(2005)]{km05}
V.~M. {Kaspi}, and M.~A. {McLaughlin}, \emph{\apjl} \textbf{618}, L41
  (2005).

\bibitem[{Pivovaroff} et~al.(2000)]{pkc00}
M.~J. {Pivovaroff}, V.~M. {Kaspi}, and F.~{Camilo}, \emph{\apj} \textbf{535},
  379 (2000).

\bibitem[{Olausen} et~al.(2010)]{okl+10}
S.~A. {Olausen}, V.~M. {Kaspi}, A.~G. {Lyne}, and M.~{Kramer}, \emph{\apj}
  (2010), in press, \eprint{arXiv:1007.1196}.

\bibitem[{McLaughlin} et~al.(2007)]{mrg+07}
M.~A. {McLaughlin}, N.~{Rea}, B.~M. {Gaensler}, S.~{Chatterjee}, F.~{Camilo},
  M.~{Kramer}, D.~R. {Lorimer}, A.~G. {Lyne}, G.~L. {Israel}, and
  A.~{Possenti}, \emph{\apj} \textbf{670}, 1307 (2007).

\bibitem[{Ng} et~al.(2008)]{nsg+08}
C.-Y. {Ng}, P.~O. {Slane}, B.~M. {Gaensler}, and J.~P. {Hughes}, \emph{\apj}
  \textbf{686}, 508 (2008).

\bibitem[{Safi-Harb} and {Kumar}(2008)]{sk08}
S.~{Safi-Harb}, and H.~S. {Kumar}, \emph{\apj} \textbf{684}, 532 (2008).

\bibitem[{Kaplan} et~al.(2009)]{kec+09}
D.~L. {Kaplan}, P.~{Esposito}, S.~{Chatterjee}, A.~{Possenti}, M.~A.
  {McLaughlin}, F.~{Camilo}, D.~{Chakrabarty}, and P.~O. {Slane}, \emph{\mnras}
  \textbf{400}, 1445 (2009).

\bibitem[{Gonzalez} et~al.(2004)]{gkl+04}
M.~E. {Gonzalez}, V.~M. {Kaspi}, A.~G. {Lyne}, and M.~J. {Pivovaroff},
  \emph{\apjl} \textbf{610}, L37 (2004).

\bibitem[{Zhu} et~al.(2009)]{zkg+09}
W.~{Zhu}, V.~M. {Kaspi}, M.~E. {Gonzalez}, and A.~G. {Lyne}, \emph{\apj}
  \textbf{704}, 1321 (2009).

\bibitem[{Leahy} and {Tian}(2008)]{lt08}
D.~A. {Leahy}, and W.~W. {Tian}, \emph{\aap} \textbf{480}, L25 (2008).

\bibitem[{Caswell} et~al.(2004)]{cmc04}
J.~L. {Caswell}, N.~M. {McClure-Griffiths}, and M.~C.~M. {Cheung},
  \emph{\mnras} \textbf{352}, 1405 (2004).

\bibitem[{Zhu} et~al.(2010)]{zkm+10}
W. {Zhu}, et~al.\ \emph{\apj} (2010), submitted.

\bibitem[{Espinoza} et~al.(2010)]{elk+10}
C.~M. {Espinoza}, A.~G. {Lyne}, M.~{Kramer}, R.~N. {Manchester}, and V.~M.
  {Kaspi}, \emph{Nature}  (2010), submitted.

\bibitem[{McLaughlin} et~al.(2006)]{mll+06}
M.~A. {McLaughlin}, A.~G. {Lyne}, D.~R. {Lorimer}, M.~{Kramer}, A.~J.
  {Faulkner}, R.~N. {Manchester}, J.~M. {Cordes}, F.~{Camilo}, A.~{Possenti},
  I.~H. {Stairs}, G.~{Hobbs}, N.~{D'Amico}, M.~{Burgay}, and J.~T. {O'Brien},
  \emph{\nat} \textbf{439}, 817 (2006).

\bibitem[{Lyne} et~al.(2009)]{lmk+09}
A.~G. {Lyne}, M.~A. {McLaughlin}, E.~F. {Keane}, M.~{Kramer}, C.~M. {Espinoza},
  B.~W. {Stappers}, N.~T. {Palliyaguru}, and J.~{Miller}, \emph{\mnras}
  \textbf{400}, 1439 (2009).

\bibitem[{Rea} et~al.(2009)]{rmg+09}
N.~{Rea}, M.~A. {McLaughlin}, B.~M. {Gaensler}, P.~O. {Slane}, L.~{Stella},
  S.~P. {Reynolds}, M.~{Burgay}, G.~L. {Israel}, A.~{Possenti}, and
  S.~{Chatterjee}, \emph{\apjl} \textbf{703}, L41 (2009).

\bibitem[{Archibald} et~al.(2008)]{akl+08}
A.~M. {Archibald}, V.~M. {Kaspi}, M.~A. {Livingstone}, and M.~A. {McLaughlin},
  \emph{\apj} \textbf{688}, 550 (2008).

\bibitem[{Helfand} et~al.(2003)]{hcg03}
D.~J. {Helfand}, B.~F. {Collins}, and E.~V. {Gotthelf}, \emph{\apj}
  \textbf{582}, 783 (2003).

\bibitem[{Livingstone} et~al.(2006)]{lkg+06}
M.~A. {Livingstone}, V.~M. {Kaspi}, E.~V. {Gotthelf}, and L.~{Kuiper},
  \emph{\apj} \textbf{647}, 1286 (2006).

\bibitem[{Gavriil} et~al.(2008)]{ggg+08}
F.~P. {Gavriil}, M.~E. {Gonzalez}, E.~V. {Gotthelf}, V.~M. {Kaspi}, M.~A.
  {Livingstone}, and P.~M. {Woods}, \emph{Science} \textbf{319}, 1802 (2008).

\bibitem[{Kumar} and {Safi-Harb}(2008)]{ks08}
H.~S. {Kumar}, and S.~{Safi-Harb}, \emph{\apjl} \textbf{678}, L43 (2008).

\bibitem[{Kuiper} and {Hermsen}(2009)]{kh09}
L.~{Kuiper}, and W.~{Hermsen}, \emph{\aap} \textbf{501}, 1031 (2009).

\bibitem[{Livingstone} et~al.(2010{\natexlab{a}})]{lkg10}
M.~A. {Livingstone}, V.~M. {Kaspi}, and F.~P. {Gavriil}, \emph{\apj}
  \textbf{710}, 1710 (2010{\natexlab{a}}).

\bibitem[{Livingstone} et~al.(2010{\natexlab{b}})]{lnk+10}
M.~A. {Livingstone}, C.-Y. {Ng}, V.~M. {Kaspi}, F.~P. {Gavriil}, and E.~V.
  {Gotthelf}, \emph{\apj}  (2010{\natexlab{b}}), in press,
  \eprint{arXiv:1007.2829}.

\bibitem[{Camilo} et~al.(2000)]{ckl+00}
F.~{Camilo}, V.~M. {Kaspi}, A.~G. {Lyne}, R.~N. {Manchester}, J.~F. {Bell},
  N.~{D'Amico}, N.~P.~F. {McKay}, and F.~{Crawford}, \emph{\apj} \textbf{541},
  367 (2000).

\bibitem[{Gonzalez} et~al.(2005)]{gkc+05}
M.~E. {Gonzalez}, V.~M. {Kaspi}, F.~{Camilo}, B.~M. {Gaensler}, and M.~J.
  {Pivovaroff}, \emph{\apj} \textbf{630}, 489 (2005).

\bibitem[{Gonzalez} and {Safi-Harb}(2003)]{gs03}
M.~{Gonzalez}, and S.~{Safi-Harb}, \emph{\apjl} \textbf{591}, L143 (2003).

\bibitem[{Weltevrede} et~al.(2010)]{wje10}
P.~{Weltevrede}, S.~{Johnston}, and C.~M. {Espinoza}, \emph{\apj}  (2010),
  submitted, \eprint{arXiv:1010.0857}.

\bibitem[{Gavriil} et~al.(2002)]{gkw02}
F.~P. {Gavriil}, V.~M. {Kaspi}, and P.~M. {Woods}, \emph{\nat} \textbf{419},
  142 (2002).

\bibitem[{Zhang} and {Harding}(2000)]{zh00}
B.~{Zhang}, and A.~K. {Harding}, \emph{\apjl} \textbf{535}, L51 (2000).

\bibitem[{Harding} et~al.(1999)]{hck99}
A.~K. {Harding}, I.~{Contopoulos}, and D.~{Kazanas}, \emph{\apjl} \textbf{525},
  L125 (1999).

\bibitem[{Alpar} et~al.(2001)]{aay01}
M.~A. {Alpar}, A.~{Ankay}, and E.~{Yazgan}, \emph{\apjl} \textbf{557}, L61
  (2001).

\bibitem[{Ek{\c s}{\.I}} and {Alpar}(2005)]{ea05}
K.~Y. {Ek{\c s}{\.I}}, and M.~A. {Alpar}, \emph{\apj} \textbf{620}, 390
  (2005).

\bibitem[{Pons} et~al.(2007)]{plm+07}
J.~A. {Pons}, B.~{Link}, J.~A. {Miralles}, and U.~{Geppert}, \emph{PRL} \textbf{98}, 071101 (2007).

\bibitem[{Aguilera} et~al.(2008{\natexlab{a}})]{apm08a}
D.~N. {Aguilera}, J.~A. {Pons}, and J.~A. {Miralles}, \emph{\apjl}
  \textbf{673}, L167 (2008{\natexlab{a}}).

\bibitem[{Aguilera} et~al.(2008{\natexlab{b}})]{apm08b}
D.~N. {Aguilera}, J.~A. {Pons}, and J.~A. {Miralles}, \emph{\aap} \textbf{486},
  255 (2008{\natexlab{b}}).

\bibitem[{Pons} et~al.(2009)]{pmg09}
J.~A. {Pons}, J.~A. {Miralles}, and U.~{Geppert}, \emph{\aap} \textbf{496},
  207 (2009).

\bibitem[{Geppert} et~al.(2004)]{gkp04}
U.~{Geppert}, M.~{K{\"u}ker}, and D.~{Page}, \emph{\aap} \textbf{426}, 267
  (2004).

\bibitem[{Geppert} et~al.(2006)]{gkp06}
U.~{Geppert}, M.~{K{\"u}ker}, and D.~{Page}, \emph{\aap} \textbf{457}, 937
  (2006).

\bibitem[{Dib} et~al.(2009)]{dkg09}
R.~{Dib}, V.~M. {Kaspi}, and F.~P. {Gavriil}, \emph{\apj} \textbf{702},
  614 (2009).

\end{thebibliography}

\end{document}